\documentclass[11pt,oneside]{article}
\usepackage{setspace,graphicx,amssymb,amsmath,latexsym,amsfonts,amscd,amsthm,multirow,ctable,mathdots,caption,array,diagbox,mathtools}
\usepackage{tabularx,cite,mathrsfs}
\usepackage{authblk}

\usepackage{fullpage}
\usepackage{stmaryrd}
\usepackage{rotating}

\usepackage{hyperref}

\newcommand{\bea}{\begin{eqnarray}}
\newcommand{\eea}{\end{eqnarray}}
\newcommand{\bee}{\begin{eqnarray*}}
\newcommand{\eee}{\end{eqnarray*}}
\newcommand{\al}{\begin{align*}}
\newcommand{\eal}{\end{align*}}
\newcommand{\be}{\begin{equation}}
\newcommand{\ee}{\end{equation}}

\newcommand{\bem}{\begin{pmatrix}}
\newcommand{\eem}{\end{pmatrix}}

\def\a{\alpha}
\def\b{\beta}

\def\e{\epsilon}
\def\f{\phi}

\def\j{\psi}
\def\k{\kappa}

\def\n{\nu}

\def\p{\pi}

\def\r{\rho}
\def\s{\sigma}
\def\t{\tau}
\def\th{\theta}
\def\til{\tilde}

\def\D{\Delta}

\def\Tr{{\rm Tr}}
\def\U{\Upsilon}

\newcolumntype{R}{ >{$}r <{$}}
\newcolumntype{C}{ >{$}c <{$}}
\newcolumntype{L}{ >{$}l <{$}}
\newcolumntype{F}{>{\centering\arraybackslash}m{1.5cm}}

\def\LL{\Lambda}

\newcommand{\wtl}[1]{\widetilde{#1}}

\newcommand{\RR}{{\mathbb R}}
\newcommand{\CC}{{\mathbb C}}
\newcommand{\PP}{{\mathbb P}}
\newcommand{\ZZ}{{\mathbb Z}}

\newcommand{\Aut}{\operatorname{Aut}}

\newcommand{\Span}{\operatorname{Span}}

\newcommand{\Id}{\operatorname{Id}}

\newcommand{\ad}{\operatorname{ad}}

\newcommand{\Sp}{\operatorname{\textsl{Sp}}}      
\renewcommand{\U}{\operatorname{\textsl{U}}}    
\newcommand{\SU}{\operatorname{\textsl{SU}}}    
\newcommand{\SO}{\operatorname{\textsl{SO}}}    
\newcommand{\Spin}{\operatorname{\textsl{Spin}}}    

\newcommand{\Co}{\textsl{Co}}	
\newcommand{\vsn}{V^{s\natural}} 
\newcommand{\tw}{{\rm tw}} 
\newcommand{\vv}{{\bf v}}

\theoremstyle{definition}

\theoremstyle{remark}

\numberwithin{equation}{section}

\begin{document}

\setstretch{1.4}

\title{\vspace{-65pt}
\vspace{20pt}
    \textsc{Equivariant K3 Invariants
    }
}

\author[1]{Miranda C. N. Cheng
}
\author[2]{John F. R. Duncan
}
\author[3]{Sarah M. Harrison
}
\author[4]{Shamit Kachru
}

\affil[1]{Institute of Physics and Korteweg-de Vries Institute for Mathematics\\
University of Amsterdam, Amsterdam, the Netherlands\footnote{On leave from CNRS, Paris.}}
\affil[2]{Department of Mathematics and Computer Science\\
Emory University, Atlanta, GA 30322, USA}
\affil[3]{Center for the Fundamental Laws of Nature\\ 
Harvard University, Cambridge, MA 02138, USA}
\affil[4]{Stanford Institute for Theoretical Physics, Department of Physics\\
and Theory Group, SLAC\\
Stanford University, Stanford, CA 94305, USA}

\date{}

\maketitle

\vspace{-1em}

\abstract{
In this note, we describe a connection between the enumerative geometry of curves in K3 surfaces and the chiral ring of an auxiliary superconformal field theory.
We consider the invariants calculated by Yau--Zaslow (capturing the Euler characters of the moduli spaces of D2-branes on curves of given genus), together with their refinements to carry additional quantum numbers by Katz--Klemm--Vafa (KKV), 
and Katz--Klemm--Pandharipande (KKP).  
We show that these invariants can be reproduced by studying the Ramond ground states of an auxiliary 
chiral superconformal field theory
which has recently been observed to give rise to mock modular moonshine for a variety of sporadic simple groups that are subgroups of Conway's group. 
We also study equivariant versions of these invariants.  A K3 sigma model is specified by a choice of 4-plane in 
the K3 D-brane charge lattice. 
Symmetries of K3 sigma models are naturally identified with 4-plane preserving subgroups of the Conway group, 
according to the work of Gaberdiel--Hohenegger--Volpato, and one may consider corresponding equivariant
refined K3 Gopakumar--Vafa invariants. 
The same symmetries naturally arise in the auxiliary CFT state space, affording a suggestive alternative view of the same computation.
We comment on a lift of this story to the generating function of elliptic genera of symmetric products of K3 surfaces.
}

\clearpage

\tableofcontents

\section{Introduction}
\label{Introduction}

The indices counting BPS states are among 
the most characteristic quantities arising from supersymmetric string compactification, and are moreover often amenable to exact evaluation. The computation of BPS indices 
has led to great insights in various topics, including supersymmetric gauge theories and quantum black holes.  
Mathematically, they are often directly related to quantities of interest in enumerative geometry. 
For instance, the Gromov--Witten invariants of a Calabi--Yau threefold $M$ are rational numbers arising naturally from the partition function of the topological A-model string theory on $M$. 
At the same time, they have the geometrical meaning of counting, roughly speaking, the holomorphic curves of a given homology class in $M$. 
Using M-theory, these invariants can (conjecturally) be recast into the Gopakumar--Vafa (GV) invariants \cite{GVone,GVtwo}, which are integers counting M2-brane configurations wrapping  supersymmetric cycles in $M$ of a given homology class. 
For a recent detailed and pedagogical discussion of these invariants, see \cite{Witten}.

K3 surfaces furnish a particularly interesting and tractable setup in which one can study BPS invariants from both a physical and a mathematical point of view. 
On the one hand, they are the simplest 
non-toroidal Calabi--Yau manifolds.  On the other hand, they are implicated in a host of dualities, which also extend nicely to K3-fibered Calabi--Yau manifolds of higher dimension. Moreover, it has become apparent in recent years that K3 compactifications of string theory have interesting connections with sporadic groups. This was perhaps first suggested by Mukai's result \cite{Mukai}, on the embedding of finite symplectomorphism groups of K3 surfaces in Mathieu's sporadic group $M_{23}$. But the significance 
for string theory was emphasized 
more recently by the {Mathieu moonshine} observation \cite{EOT} of Eguchi--Ooguri--Tachikawa, which connects the largest Mathieu group $M_{24}$ and the K3 elliptic genus. 

This connection has led to a surge of activity in the study of 
correspondences relating sporadic groups and  mock modular forms, and in particular led to the discovery of 
{umbral moonshine} \cite{UM,UMNL,mumcor}, whose relation to string compactification on K3 has been discussed in \cite{Harvey:2013mda,Harvey:2014cva,Cheng:2014zpa}. 
The illumination of (some aspects of) the role of sporadic groups and moonshine in string theory on K3 constitutes one of the main motivations for this note. 

The burden of this work is to argue that the enumerative invariants of K3 surfaces are in fact controlled by 
the action of 4-plane preserving subgroups of Conway's group, $\Co_0$, the automorphism group of the Leech lattice. Working over $\CC$, the 24-dimensional representation of $\Co_0$ 
may be identified with the K3 cohomology space. 
We show that this identification naturally leads to a group action on the moduli spaces of BPS objects underlying the enumerative invariants of Yau--Zaslow \cite{YauZaslow}, the refinements of these defined by Katz--Klemm--Vafa \cite{KKV}, and the (further refined) motivic stable pairs invariants of Katz--Klemm--Pandharipande \cite{KKP}.  This allows us to easily reproduce these invariants as flavored partition functions in an auxiliary theory of 24 bosons.
We codify our claims in three conjectures, which sequentially describe the symmetry of the graded space of BPS states, the 
equivariant reduced Gopakumar--Vafa invariants defined by suitable automorphisms (or autoequivalences) 
of K3 surfaces, and 
the 
equivariant 
refined reduced Gopakumar--Vafa invariants.

The $24$-dimensional representation of $\Co_0$ also arises as the chiral ring (or associated Ramond ground states) in $V^{s\natural}$, the state space of a chiral superconformal field theory which plays a starring role in moonshine for the Conway groups \cite{Duncan,DuncanMack-Crane1}.  Consequently, we will see that one can think of the K3 invariants as being given by
traces in this state space.  Beyond its role in Conway moonshine,
 $\vsn$ has recently been used to attach Jacobi forms to derived autoequivalences of K3 surfaces in \cite{DuncanK3}, and it has been argued that these Jacobi forms can reproduce twined elliptic genera defined by supersymmetry preserving automorphisms of K3 sigma models. Moreover, $\vsn$ has been shown to underly mock modular moonshine for a variety of sporadic simple groups in \cite{M5}, where Jacobi forms (of a different kind) have also appeared.

Our conjectures are motivated, stated and explained in \S\ref{sec:Symmetries and Twinings}. In advance of this, we review BPS states in K3 compactifications, and the refined enumerative invariants of K3 surfaces in \S\ref{sec:K3rev}. We review the distinguished vertex operator superalgebra $\vsn$ in \S\ref{sec:The Supernatural CFT}.
We conclude with a few comments and further open questions in \S\ref{sec:disc}. We make use of the Jacobi theta functions in what follows. We record their definitions in \S\ref{sec:theta}.

We note before preceding that, as is always the case, our present article relies upon earlier works. We mention 
\cite{Volpato,Huybrechts:2013iwa,Benjamin:2014kna,M5spin7} 
in particular,
in addition to those works already cited, 
as some of the most relevant antecedents.

\section{Refined K3 Invariants}
\label{sec:K3rev}

  To see how BPS indices of a K3 compactification can be related to certain statements about the enumerative geometric properties of K3 surfaces, first recall the following equality. 
  The generating function of the Euler characteristic of $K3^{[n]}$, the Hilbert scheme of $n$ points on an arbitrary K3 surface, is given
\cite{MR1032930} 
by G\"ottsche's formula
  \be\label{euler1}
  \sum_{n\geq 0} \chi (K3^{[n]}) q^{n-1} = \frac{1}{\D(\t)},
  \ee  
where
$q=e^{2\pi i \t}$, and $\D(\t)$ denotes the 
unique (up to scale) cuspidal modular form of weight $12$ for the modular group, 
  \be
  \D(\t) = \eta^{24}(\t) = q\prod_{k>0} (1-q^k)^{24}.
  \ee
To see how the above quantity relates to BPS indices, consider the bound states of one D4-brane and $n$ D0-branes on a K3 surface in type IIA string theory. The corresponding BPS index is given by the Euler characteristic of the relevant moduli space, which in this case is of the form $K3^{[n]}$.
  Recall that $K3^{[n]}$ 
  serves as a 
  desingularisation of the $n$-th symmetric product $K3^{(n)}:= K3^n/S_n$.
  As shown in \cite{MR1481482}, the value $\chi (K3^{[n]})$ coincides with the orbifold Euler characteristic of $K3^{(n)}$ (and directly similar statements hold for Hodge numbers). 
Using the string duality relating type IIA string theory compactified on K3 to heterotic string theory compactified on a torus, $T^4$, the  right-hand side of (\ref{euler1}) can be understood as 
the  chiral partition function of a bosonic string. This is counting the Dabholkar--Harvey states---half BPS states of the heterotic theory---with the right-moving oscillators in their ground state and arbitrary left-moving excitations.  
  
The 
relation of 
(\ref{euler1}) to K3 curve counting, discussed by  Yau--Zaslow \cite{YauZaslow}, building on the physical results of \cite{Bershadsky:1995qy,Vafa, VafaWitten}, is best understood when we go to 
another duality frame. Namely, the BPS system of one D4-brane and $n$ D0-branes on K3 is dual to the system of one D2-brane wrapping a $2$-cycle of K3 with self-intersection number $2n-2$, and this can 
be realised as a holomorphic curve of genus $n$ whenever $n\geq 0$. Hence, in the D2-brane duality frame, the BPS index is 
argued to be equal to the Euler characteristic of the (compactification of  the) moduli space of holomorphic curves of genus $n$ with a choice of flat $\U(1)$-bundle \cite{Bershadsky:1995qy}. Let us denote this moduli space by ${\cal M}_n^H$. In \cite{YauZaslow} it was shown that the contribution to $\chi({\cal M}_n^H)$ is localised on curves with genus $0$ and $n$ double (nodal) points. From the above one arrives at the 
Yau--Zaslow formula
 \be\label{YauZaslow}
 \sum_{n\geq 0} d_n q^{n-1} = \frac{1}{\D(\t)} = q^{-1}( 1  +  24 q + 324 q^2 + 3200 q^3 + \cdots),
 \ee
where $d_n$ counts the number of $\PP^1$'s with $n$ double points in a K3 surface. (See \cite{MR1682284} for a 
proof of (\ref{YauZaslow}).) 

The Euler characteristics can be viewed as the special limit of the $\chi_y$-genus, $ \chi_y(M) := \sum_{p,q}y^{p}(-1)^{q} h^{p,q}(M)$. The generating function 
of $\chi_{-y} (K3^{[n]})$ is given by 
\begin{equation}\label{eqn:chiy}
\begin{split}
\sum_{n \geq 0}
\chi_{-y} (K3^{[n]})y^{-n}q^{n-1}  &= 
q^{-1}\prod_{k>0} (1-yq^k)^{-2} (1-q^k)^{-20} (1-y^{-1}q^k)^{-2}\\
&=(-y+2-y^{-1})\frac{\eta(\tau)^6}{\theta_1^2(\tau,z)}\frac1{\Delta(\tau)}.
\end{split}
\end{equation}

Note that the above formula can also be derived as the special limit of a stringy generalisation of the G\"ottsche formula. 
The generating function 
formula (\ref{euler1}) has been extended in \cite{DVV,DMVV} from the Euler characteristic, to the elliptic genus \cite{Ochanine,WittenEGQFT,Landweber,WittenLGN2,KYY}.
Writing the Fourier expansion of the K3 elliptic genus as 
\be\label{eqn:c(4n-ell2)}
Z_{\rm EG}(\t,z;K3) = \sum_{\substack{n, \ell \in \ZZ\\n\geq0}} c(4n-\ell^2)\,q^n y^\ell = 2 y+2y^{-1}+20 + O(q),
\ee
where 
$y=e^{2\pi i z}$, 
the second quantised K3 elliptic genus is given \cite{DMVV} by the 
DMVV formula:
\be\label{hugehog}
\sum_{n\geq 0} 
Z_{\rm EG}(\t,z;K3^{[n]}) p^{n-1}  =  
	p^{-1} \prod_{\substack{r,s,t\in\ZZ\\r>0, s\geq 0}} 
	{(1-q^sy^tp^r)^{-c(4rs-t^2)}}.
\ee
Recall that the elliptic genus is 
a generalisation of Hirzebruch's $\chi_y$ genus: 
for a compact complex manifold $M$ with complex dimension $d$, we have
\be\label{chi_y_EG}
\lim_{\t\to i\infty} Z_{\rm EG}(\t,z;M) = y^{-d/2}\chi_{-y}(M).
\ee
As a result, taking $\tau\to i\infty$ in 
(\ref{hugehog}) (and then replacing $p$ with $q$), we arrive at (\ref{eqn:chiy}). 

Given the 
geometric interpretation of the generating function 
(\ref{euler1}) just discussed, 
it is natural to ask whether this one-variable refinement (\ref{eqn:chiy}) 
also admits a 
curve-counting interpretation. 
Following \cite{GVone,GVtwo}, Katz--Klemm--Vafa  proposed in \cite{KKV} that the numbers $n_{n}^r$, satisfying 
 \begin{equation}\label{eqn:KKV}
\sum_{r\geq 0} \sum_{n\geq 0} (-1)^r n_{n}^r (y^{1/2}-y^{-1/2})^{2r} q^{n-1} = 
q^{-1}\prod_{k>0} (1-yq^k)^{-2} (1-q^k)^{-20} (1-y^{-1}q^k)^{-2},
\end{equation}
encode the reduced Gromov--Witten invariants of a K3 surface in the following way. 
Given a (smooth projective) K3 surface $X$, and a primitive class $\a \in {\rm Pic}(X)$, the Gromov--Witten potential reads
 \be
 F_\a (g_s,x)= \sum_{r\geq 0} \sum_{k\geq 1} R^r_{k\a} \,g_s^{2r-2} e^{x\cdot k\a},
 \ee
 where $R^r_{\b}$ is the reduced Gromov--Witten invariant of genus $r$ and curve class $\b$. 
It can be re-expressed as
\be
F_\a (g_s,x)= \sum_{r\geq 0} \sum_{k \geq 1} {\tilde n}_{k\a}^r \,g_s^{2r-2}\sum_{d>0} \frac{1}{d} 
\left( \frac{\sin(dg_s/2)}{g_s/2}\right)^{2r-2}  e^{x\cdot d\k\a},
\ee
where ${\tilde n}_{\b}^r=n_{n}^r$ for all curve classes $\b$ with self-intersection number $\b\cdot\b=2n-2$.
Note that by specialising (\ref{eqn:KKV}) to $z=0$ we recover the Yau--Zaslow formula (\ref{YauZaslow}) with $d_n = n_n^{r=0}$.
The KKV conjecture relating (\ref{eqn:KKV}) to Gromov--Witten invariants has recently been proven in \cite{Pandharipande:2014qoa}. (See also \cite{2014arXiv1411.0896P}.)

After refining from the Euler characteristic (\ref{euler1}) to the $\chi_y$ genus (\ref{eqn:chiy}), the next obvious refinement is the generating function \cite{MR1481482} for the Hodge polynomials of $K3^{[n]}$,
\be\label{eqn:Hodge}
\begin{split}
&\sum_{n \geq 0}
\chi_{\rm Hodge} (K3^{[n]}) 
q^{n-1} 
\\ 
&={q^{-1} {\prod_{k>0} (1-uy q^k)^{-1}(1-u^{-1} yq^k)^{-1}(1-q^k)^{-20}(1-uy^{-1} q^k)^{-1} (1-u^{-1}y^{-1} q^k)^{-1}  } }\\
&=(u-y-y^{-1}+u^{-1})\frac{\eta(\tau)^6}{\theta_1(\tau,z+w)\theta_1(\tau,z-w)}\frac1{\Delta(\tau)},
\end{split}
\ee
where $u=e^{2\pi i w}$.
Here, following \cite{KKP}, we define 
\begin{gather}
\chi_{\rm Hodge}(M):= u^{-d/2}y^{-d/2}\sum_{p,q}  (-u)^{q}(-y)^{p}  h^{p,q}(M)
\end{gather} 
for $M$ a K\"ahler manifold of complex dimension $d$. 
As before, $K3^{[n]}$ denotes the Hilbert scheme of $n$ points on an arbitrary K3 surface.

To discuss the enumerative geometric interpretation of (\ref{eqn:Hodge}) 
it will be helpful to first recall the physical origin of the (reduced refined) Gopakumar--Vafa invariants.   
Given a Calabi--Yau threefold $M$, and a homology class $\b\in H_2(M,\ZZ)$, the refined Gopakumar--Vafa invariant $\til N_\beta^{j_L,j_R}$ is defined as the BPS index counting the BPS states of M2-branes wrapping a $2$-cycle of class $\b$, 
with spin quantum numbers $(j_L,j_R)$ under the little group $\SO(4)\simeq \SU(2)_L\times \SU(2)_R$ of a massive particle in 5-dimensional Lorentzian space-time. 
More directly, consider  M-theory on $M\times S^1 \times_{\vec\e} TN $ where $TN$ denotes the Taub-NUT space, and the subscript $\vec\e=(\e_1,\e_2)$ indicates a twist by the element $(e^{i\e_1},e^{i\e_2})$ of the $\U(1)\times \U(1)$ subgroup of the Taub-NUT isometry group, $\U(1)\times \SO(3)$. 
Then, in particular, the geometry is locally given by $\CC^2$ near the tip of the Taub-NUT factor, and writing $(z_1,z_2)$ for local coordinates near the tip, the twist along the $S^1$ is given by 
 $$
 z_1 \to e^{i\e_1} z_1 , ~z_2 \to e^{i\e_2} z_2.  
 $$
The  BPS partition function of this M-theory compactification, in the presence of suitable three-form fluxes, is given by 
\be\label{def:ZBPS}
\begin{split} 
&Z_{\rm BPS}=\\
&\prod_{\b,j_L,j_R} \prod_{m_L=-j_L}^{j_L}\prod_{m_R=-j_R}^{j_R} \prod_{n_1,n_2\geq 0}(1-e^{-\e_1(m_++n_1+\frac12)}e^{\e_2(m_-+n_2+\frac12)}e^{-x\cdot\beta})^{(-1)^{2j_L+2j_R}\til N_\beta^{j_L,j_R}}
\end{split}
\ee
according to \cite{Dijkgraaf:2006um,Denef:2007vg,Aganagic:2009kf}. 
Here, $m_{\pm} := m_L \pm m_R$. The $m_L$ and $m_R$ specify the spin content of a given
BPS multiplet under the $\SU(2)_L \times \SU(2)_R$ little group of M-theory compactification to 5d.
The $n_1$ and $n_2$ 
control the orbital helicities of the BPS states in the two $2$-planes of the
local approximate ${\mathbb C}^2$ near the center of the Taub-NUT factor.  The formula (\ref{def:ZBPS}) can roughly be understood as an
enumeration of 
states with various
``chemical potentials'' for angular momentum and helicity. The $(-1)^{2j_L + 2j_R}$ factor in 
the exponent gives the
appropriate Boltzmann factor for bosonic and fermionic states. The $\tilde N^{j_L,j_R}_{\beta}$ simply count the 
number of 
multiplets at each set of quantum numbers.

In order for the background to have unbroken supersymmetry for generic $\e_1$ and $\e_2$, one should include an extra $\U(1)_R$ twist along the Calabi--Yau threefold $M$. 
This extra $\U(1)$ symmetry is available for $M = K3\times T^2$, but is unavailable for compact threefolds with $\SU(3)$ holonomy. In the latter case 
one must impose $\e_1+\e_2=0$ in order to preserve supersymmetry, and hence one is led to the usual Gopakumar--Vafa invariants $ \til n_\b^{j}$, 
satisfying
\be\label{rel_refined_unrefinedGV}
\sum_{j\in \ZZ_{\geq 0}} (y^{1/2}-y^{-1/2})^{2j} (-1)^{j} \til n_\b^{j} = \sum_{j_R,j_L\in \frac{1}{2}\ZZ_{\geq0}} (-1)^{2(j_R+j_L)} (2j_R+1) \til N_\beta^{j_L,j_R} 
[j_L]_y,
\ee
where 
$[j]_x := x^{-2j}+x^{-2j+2}+\dots + x^{2j}$ for $j\in \frac12\ZZ$.
 In the cases where it is possible to allow for independent $\e_1$ and $\e_2$, another interesting limit is the so-called Nekrasov--Shatashvili (NS) limit \cite{Nekrasov:2009uh,Nekrasov:2009rc}, where we take one of the $\e$'s to zero. See for instance \cite{Aganagic:2011mi,Hatsuda:2013oxa,Krefl:2013bsa,Grassi:2014zfa,Huang:2014nwa}, for more on the physical meaning and significance of this limit. 
 
To define the refined Gopakumar--Vafa invariants on K3 surfaces, we consider Calabi-Yau threefolds
of the form $M=K3\times T^2$ and restrict attention to the curve classes $\beta$ that lie in  $H_2(K3,\ZZ)$. (For a general Calabi--Yau threefold certain subtleties involving wall-crossing may arise. We refer the reader to \cite{Denef:2007vg} for more on this.)
In other words, given a K3 surface $X$ we consider the BPS index of the M-theory lift of the type IIA configuration consisting of a D6-brane wrapping $X\times T^2$, and D2-brane wrapping a curve inside $X$ with curve class $\beta$.
Note that, in order to get a non-vanishing answer for such invariants we need to consider the so-called reduced refined Gopakumar--Vafa invariants, obtained by discarding certain extra $\SU(2)$-multiplets originating from the $\U(1)\times \U(1)$ symmetry of the extra $T^2$. This is explained in detail in \cite{KKV}. Henceforth, when we speak of (refined) K3 Gopakumar--Vafa invariants, we will mean the reduced (refined) Gopakumar--Vafa invariants associated in this way to K3 surfaces.

Note that a choice of complex structure 
and 
complexified K\"ahler class $B+iJ$ for $X$ is implicit in the above setting. 
This data determines a positive-definite $4$-dimensional subspace $\Pi$ of the real cohomology space 
$\widetilde{H}(K3,\ZZ)\otimes_\ZZ \RR$, 
which will play an important role in what follows. (Cf. (\ref{eqn:defnPi}) and \S\ref{sec:Symmetries and Twinings}.)

The refined Gopakumar--Vafa invariants $\til N_\beta^{j_L,j_R}$ admit a more mathematical definition  
in terms of certain refinements  \cite{Choi:2012jz} of the Pandharipande--Thomas invariants. These, in turn, are defined \cite{PTinv,PTBPS} in terms of  the moduli space of stable pairs on a threefold $M$ that capture the 
degeneracies of D6-D2-D0 bound states. 
Alternatively,  one can define the refined Gopakumar--Vafa invariants via a
suitably-defined moduli space of semi-stable coherent sheaves that are pure of dimension $1$ on $M$---the so-called M2-brane moduli space---equipped with 
a relative Lefschetz action of $\SU(2)_L\times \SU(2)_R$ on its cohomology.
This can be deduced from \cite{MR1849482}.
In the context of the present paper we proceed as in \cite{KKP,Pandharipande:2014qoa}, considering K3-fibered Calabi--Yau threefolds $M$, and restricting attention to the fibre curve classes $\beta\in H_2(M,\ZZ)$.

Adopting the above two definitions respectively, 
 it was conjectured in \cite{KKP}, and its appendix by Thomas, respectively,
that for $X$ a K3 surface we have $\til N_\beta^{j_L,j_R}=  N_n^{j_L,j_R}$ for all $\b\in H_2(X,\ZZ)$ with self-intersection $\b\cdot\b=2n-2$, where the $N_n^{j_L,j_R}$ are defined by requiring that
\begin{gather}
	\begin{split}
\label{refined}
&\sum_{n\geq 0} \sum_{\substack{j_L,j_R \in \frac{1}{2}\ZZ\\j_L,j_R\geq 0}} N_n^{j_L,j_R} [j_L]_y [j_R]_u q^{n-1} \\&=
q^{-1}\prod_{k>0} {(1-uyq^k)^{-1}(1-u^{-1}yq^k)^{-1}(1-q^k)^{-20}(1-uy^{-1}q^k)^{-1}(1-u^{-1}y^{-1}q^k)^{-1}}.
\end{split}
\end{gather} 
See \cite{Huang:2013yta} for related earlier work. Note that we have omitted the factor $(-1)^{2(j_L+j_R)}$ since in the K3 case we have $N_n^{j_L,j_R}=0$ unless $j_L+j_R\in \ZZ$. 

Given the 
renewed interest in the relationship between BPS states in K3 compactifications and
sporadic groups arising from the recent 
Mathieu moonshine observation \cite{EOT}, 
the authors of \cite{KKP} raised a natural question: are these refined invariants $N_n^{j_L,j_R}$ related to sporadic groups as well? (See \S5.5 of \cite{KKP}.) In this note we 
provide 
an affirmative answer to this question, 
by positing concrete falsifiable 
conjectures which state, in essence, that equivariant versions of the refined invariants $N_n^{j_L,j_R}$, defined by 
supersymmetry-preserving automorphisms of the 
underlying D-brane moduli spaces, are naturally 
related to 4-plane preserving subgroups of the sporadic group $\Co_0$.  As we further discuss, they can also be
recovered
from the geometry of ground states of the chiral superconformal field theory $\vsn$. In particular, the $N_n^{j_L,j_R}$ are related to the sporadic group $\Co_1$, and some of its sporadic subgroups, via the action of these groups on $\vsn$.

In the next section we 
give a brief review of the relevant features of 
$\vsn$.

\section{Chiral Conformal Field Theory}
\label{sec:The Supernatural CFT}

In 
this section we  review the 
chiral superconformal field theory (i.e. vertex operator superalgebra)
$V^{s\natural}$,
which 
plays a central role in various 
moonshine phenomena relating (mock) modular forms to sporadic simple groups.  

Consider the $2$-dimensional theory of
$24$ chiral free fermions $\psi_i$, orbifolded by the ${\mathbb Z}/2$ symmetry
\begin{equation}
\psi_i \mapsto -\psi_i.
\end{equation}
This model turns out to be equivalent (i.e., isomorphic as a vertex operator superalgebra, cf. \cite{Duncan}) to a superconformal field theory first discussed in \cite{FLM}, 
described there in terms of a supersymmetric extension of the $E_8$ current
algebra. Write $\Co_0$ for Conway's group, being the automorphism group of the Leech lattice \cite{MR0237634,MR0248216}. The quotient $\Co_1:=\Co_0/\{\pm \Id\}$ is Conway's largest sporadic simple group (cf. \cite{ATLAS}).
As observed in \cite{FLM}, the Neveu--Schwarz sector partition function 
\be\label{eqn:vsnzns}
\begin{split}
Z_{\rm NS}(\t)&:=\Tr(q^{L(0)-\frac{c}{24}}|\vsn)\\
&\;= \frac{1}{2} \frac{1}{\eta^{12}(\tau)} \sum_{i=2}^4 \theta_i^{12}(\tau,0)= q^{-1/2} +276 q^{1/2} + 2048q + 11202q^{3/2}+
\dots
\end{split}
\ee
suggests a relation to 
$\Co_1$,
as the numbers $276$, $2048$, $11202$, are 
dimensions of 
non-trivial representations of this group. Interestingly, and importantly, the coefficients of integer powers of $q$ in (\ref{eqn:vsnzns}) can also be interpreted as dimensions of faithful representations of $\Co_0$.

These observations initiate a theory parallel to---but not disjoint with---monstrous moonshine \cite{CN,FLMbook,Borcherds,DuncanGriffinOno}, in which supersymmetric string theory 
replaces the bosonic string, and 
Conway's group replaces the monster. Certain details of this Conway moonshine have been elucidated in \cite{Duncan,DuncanMack-Crane1}.
The Conway moonshine model has also played an important role in attempts to find a dual for pure supergravity in
$AdS_3$ \cite{Wittentwo}.
Recent, comprehensive discussions focusing on the features most relevant for us can
be found in \cite{DuncanMack-Crane1,DuncanK3,Benjamin:2014kna,M5spin7,M5}.

As alluded to above, this theory enjoys (hidden) ${\cal N}=1$ supersymmetry. To be precise, there is a linear combination of the
spin-3/2 twist fields, intertwining the 
Neveu--Schwarz and Ramond sectors of the ${\mathbb Z}/2$ orbifold,
which satisfy the OPEs of an ${\cal N}=1$ supercurrent with central charge $c=12$.  
The subgroup of $\Spin(24)$ that preserves this 
${\cal N}=1$ supercurrent 
turns out to be 
Conway's group $\Co_0$ \cite{Duncan}. 
We write $\vsn$ for the vertex operator superalgebra 
which constitutes the Neveu--Schwarz sector of this 
model, and we write $\vsn_\tw$ for the unique canonically-twisted $\vsn$-module for $\vsn$ (cf. \cite{Duncan,DuncanMack-Crane1}), which is the Ramond sector of the model. Note that $\vsn$ and $\vsn_\tw$ are both faithful as modules for $\Co_0$. The $24$ fermions $\psi_i$ constitute the ground states of $\vsn_\tw$, and transform according to the $24$-dimensional Leech lattice representation of $\Co_0$. 
We write ${\bf 24}$ for the $\Co_0$ module afforded by the Ramond sector ground states in $\vsn_\tw$. If we let $\tau_\tw$ denote the vector 
that generates the aforementioned ${\cal N}=1$ supercurrent, then $\tau_\tw\in\vsn_\tw$ is the unique (up to scale) spin-$3/2$ vector in $\vsn\oplus\vsn_\tw$ that is fixed by the $\Co_0$-action.

It is natural to use linear combinations 
of the fermions $\psi_i$ 
to form a current algebra. This is the basic idea underlying the analyses of \cite{DuncanK3,M5,M5spin7}. 
For example, if $a$ is an arbitrary non-zero linear combination of the $\psi_i$, then the natural action of $a$ on $\tau_\tw$ determines an ${\cal N}=1$ supercurrent $\tau_a:=a(0)\tau_\tw$ 
in the Neveu--Schwarz sector, $\vsn$. As is discussed in detail in \cite{M5spin7}, there is a choice of $a$ such that the subgroup of $\Co_0$ that fixes $\tau_a$ is precisely the largest sporadic 
Mathieu group, $M_{24}$, and this fact can be used to attach mock modular forms of weight $1/2$ to the conjugacy classes of $M_{24}$ (although these forms are different from those arising from the Mathieu moonshine observation \cite{EOT} of Eguchi--Ooguri--Tachikawa).

In \cite{DuncanK3} stability conditions on K3 surfaces are used to identify orthogonal pairs of 
Ramond sector ground states, $a_X^\pm,a_Z^\pm\in{\bf 24}$. 
Such a pair 
determines a supersymmetric K3 sigma model, according to the description \cite{MR1479699,MR1416354,NahWen} in terms of positive-definite $4$-planes in the real K3 cohomology space, 
and can also be used to equip $\vsn$ with an action of the ${\cal N}=4$ superalgebra at central charge $c=6$, as we will 
demonstrate momentarily. Since it will be important in what follows, we briefly review the construction of $a_X^\pm,a_Z^\pm$, and the corresponding $4$-plane $\Pi=\Pi_{X,Z}$ now, referring to \cite{DuncanK3} for a fuller discussion. 

To begin, let $X$ be a 
(smooth projective) K3 surface, 
and choose vectors $a_X^-\in H^{2,0}(X)$ and $a_X^+\in H^{0,2}(X)$ such that $\langle a_X^\pm,a_X^\mp\rangle=1$, where $\langle\cdot\,,\cdot\rangle$ denotes the Mukai pairing (cf., e.g. \cite{DuncanK3} or \cite{Huybrechts:2013iwa}) on the K3 cohomology lattice $\wtl{H}(X,\ZZ):=\bigoplus H^n(X,\ZZ)$, extended linearly to the complex envelope, $\wtl{H}(X,\ZZ)\otimes_\ZZ\CC=\bigoplus H^{p,q}(X)$. (Note that $H^{2,0}(X)$ and $H^{0,2}(X)$ are isotropic with respect to $\langle\cdot\,,\cdot\rangle$.) Next, pick a stability condition in Bridgeland's distinguished component \cite{BridgelandK3} of the space of stability conditions on $X$, write $Z$ for the corresponding central charge, which we may regard as an element of $H^{0,0}(X)\oplus H^{1,1}(X)\oplus H^{2,2}(X)$, and choose isotropic $a_Z^\pm\in\CC Z\oplus \CC\bar{Z}$ such that $\langle a_Z^\pm,a_Z^\mp\rangle=1$. Then
\begin{gather}\label{eqn:defnPi}
	\Pi=\Pi_{X,Z}:=\Span\left\{a_X^\pm,a_Z^\pm\right\}\cap \wtl{H}(X,\ZZ)\otimes_\ZZ\RR
\end{gather}
is a maximal positive-definite subspace of $\wtl{H}(X,\ZZ)\otimes_\ZZ\RR\simeq \RR^{4,20}$ such that the negative-definite lattice $\Pi^\perp\cap\wtl{H}(X,\ZZ)$ has no vectors $\lambda$ with $\langle\lambda,\lambda\rangle=-2$ (cf. \S4 of \cite{DuncanK3}). As such, $\Pi$ determines a supersymmetric non-linear sigma model on $X$, according to \cite{MR1479699,MR1416354}. (Cf. also the discussion in \cite{Volpato}.) Note that all such $\Pi$ arise as $\Pi_{X,Z}$ for some $X$ and $Z$ as above, according to the main result of \cite{Huybrechts:2013iwa}. If $B+iJ$ is a complexified K\"ahler class for $X$, as in \S\ref{sec:K3rev}, then the corresponding $4$-space is $\Pi=\Pi_{X,Z}$ where $Z=\exp(B+iJ)$.

In order to identify 
$a_X^\pm$ and $a_Z^\pm$ as elements of ${\bf 24}$, first consider the group 
of 
orthogonal transformations of $\wtl{H}(X,\ZZ)$ that act trivially on $\Pi$, which we denote 
\begin{gather}\label{eqn:defnGPi}
	G_\Pi:=\left\{g\in\Aut(\wtl{H}(X,\ZZ))\mid g|_{\Pi}=\Id_\Pi\right\}.
\end{gather}
Following \cite{Volpato} we regard $G_\Pi$ as the group of supersymmetry preserving automorphisms of the sigma model $\Pi$. 
Next, let $\Gamma_\Pi$ be 
the orthogonal complement in $\wtl{H}(X,\ZZ)$ of its sublattice of $G_\Pi$-fixed points, $\wtl{H}(X,\ZZ)^{G_\Pi}$. 
Then $\Gamma_\Pi$ is acted on faithfully by $G_\Pi$, and is negative-definite 
since $\Pi<\wtl{H}(X,\ZZ)^{G_\Pi}\otimes_\ZZ\RR$. 
According to \S\S B.1-2 of \cite{Volpato}, there is an isometric embedding $\Gamma_\Pi\hookrightarrow\LL(-1)$, where $\LL(-1)$ denotes the negative-definite Leech lattice. (Cf. also \S2.2 of \cite{Huybrechts:2013iwa}.) We identify $a_X^\pm$ and $a_Z^\pm$ as the 
Ramond sector ground states 
of our model
by choosing an isometry 
\begin{gather}\label{eqn:isomHXLL}
\wtl{H}(X,\ZZ)\otimes_\ZZ\CC\xrightarrow{\sim} {\bf 24}
\end{gather}
that 
maps 
$\Gamma_\Pi$ 
into 
the unique copy of $\LL(-1)$ in ${\bf 24}$ that is invariant under the action of $\Co_0$. Note that $g\in G_\Pi$ acts naturally on $\LL(-1)$, fixing the vectors orthogonal to $\Gamma_\Pi$. That is, the isometry (\ref{eqn:isomHXLL}) identifies $G_\Pi$ with a subgroup of $\Co_0$ that 
fixes the ground states $a_X^\pm$ and $a_Z^\pm$.

The vertex operator superalgebra $\vsn$ inherits an
${\cal N}=4$ superconformal structure with central charge $c=6$ 
from $a_X^\pm,a_Z^\pm\in{\bf 24}$, according to the following construction. 
Using the notation of \cite{DuncanMack-Crane1,DuncanK3}, define vectors in $\vsn$ by setting
\begin{gather}
\begin{split}\label{eqn:N4c6gens}
\jmath^3&:=\frac14(a_X^-(-\tfrac12)a_X^+(-\tfrac12)\vv+a_Z^-(-\tfrac12)a_Z^+(-\tfrac12)\vv),\\
\jmath^\pm&:=\frac i2a_X^\pm(-\tfrac12)a_Z^\pm(-\tfrac12)\vv,\\
\tau_1^\pm&:=\sqrt{2}\left(a_X^-(0)\pm a_X^+(0)\pm a_X^\mp(0)a_Z^-(0)a_Z^+(0)\right)\tau_\tw,\\
\tau_2^\pm&:=\mp i\sqrt{2}\left(a_Z^-(0)\pm a_Z^+(0)\pm a_Z^\mp(0)a_X^-(0)a_X^+(0)\right)\tau_\tw,
\end{split}
\end{gather}
where $\vv$ is the vacuum of $\vsn$.
Then $\jmath^3$ and the $\jmath^\pm$ generate an $\SU(2)$ current algebra, and the pairs $\{\tau^-_1,\tau^-_2\}$ and $\{\tau^+_1,\tau^+_2\}$ each span a copy of the natural $2$-dimensional representation of the correpsonding $\SU(2)$.
Write $G^\pm_j(z)$ for the spin-$3/2$ fields corresponding to the $\tau_j^\pm$, and 
define $J^\pm(z)$ and $J^3(z)$ similarly in terms of the $\jmath^\pm$ and $\jmath^3$. Then a routine calculation verifies the OPEs
\begin{gather}
\begin{split}\label{eqn:N4c6OPEs}
G_j^\pm(z)G^\mp_j(w)&\sim \frac{4}{(z-w)^3}\pm \frac{4J^3(w)}{(z-w)^2}+
\frac{2T(w)}{(z-w)}\pm\frac{2\partial_w J^3(z)}{(z-w)}
,\\
G_1^\pm(z)G_2^\pm(w)&\sim\pm\frac{4J^\pm(w)}{(z-w)^2}\pm\frac{2\partial_wJ^\pm(w)}{(z-w)},\\
G_j^\pm(z)G_j^\pm(w)&\sim G_1^\pm(z)G_2^\mp(w)\sim 0,
\end{split}
\end{gather}
for $j\in \{1,2\}$, where $T(z)$ is a spin-$2$ current generating the Virasoro algebra at $c=6$. 
That is to say, the $G^\pm _j(z)$ generate an action of the (small) ${\cal N}=4$ superconformal algebra at $c=6$ (cf. \cite{Eguchi1987}) on $\vsn$. Note that $T(z)$ is not the stress-energy tensor of $\vsn$, which we denote $L(z)$, and which has central charge $c=12$. More details on (\ref{eqn:N4c6gens}) and (\ref{eqn:N4c6OPEs}) will appear in \cite{CreDunRie}.

Our final objective in this section is to discuss some 
partition functions that arise naturally from $\vsn$, when equipped with one of the aforementioned superconformal structures.
We note, to begin, that the 
Ramond sector index 
is simply a constant,
\begin{gather}
\label{Rpart}
\begin{split}
Z^{s\natural}(\tau) &:= {\rm Tr}((-1)^F q^{L(0) - \frac{c}{24}}|\vsn_{\rm tw})\\
&\;= \frac{1}{2} \frac{1}{\eta^{12}(\tau)} \sum_{i=2}^4 (-1)^{i+1} \theta_i^{12}(\tau,0)=24.
\end{split}
\end{gather}
We next choose a $\U(1)$ current 
and consider the corresponding $\U(1)$-graded Ramond sector index. 
Taking $J(z):=2J^3(z)$, the $\U(1)$-graded index works out \cite{DuncanK3} to be
\begin{gather}
\label{Zgraded-DM}
\begin{split}
Z^{s\natural}(\tau,z) &:= {\rm Tr}((-1)^F q^{L(0) - \frac{c}{24}} y^{J(0)}|\vsn_\tw)\\
&\;= \frac{1}{2} \frac{1}{\eta^{12}(\tau)} \sum_{i=2}^4 (-1)^{i+1} \theta_i^2(\tau,z) \theta_i^{10}(\tau,0),
\end{split}
\end{gather}
which is a weak Jacobi form of weight $0$ and index $1$ satisfying $Z^{s\natural}(\tau,0)=Z^{s\natural}(\tau)=24$. That is, 
we have recovered the K3 elliptic genus, 
\begin{equation}\label{eqn:vsnEGK3}
Z^{s\natural}(\tau,z) =  Z_{\rm EG}(\t,z;K3).
\end{equation}

Evidence is presented in \cite{DuncanK3} that the coincidence (\ref{eqn:vsnEGK3}) is not an accident, but rather reflects a deep relationship between $\vsn$ and K3 surface geometry. To explain this, 
let $g\in G_\Pi$ 
and consider the 
$g$-twined $\U(1)$-graded Ramond sector index,
\begin{gather}
\label{Zgraded-DM-gtwined}
Z^{s\natural}_g(\tau,z) 
:= {\rm Tr}(g (-1)^F q^{L(0) - \frac{c}{24}} y^{J(0)}|\vsn_\tw).
\end{gather}
We may compare 
(\ref{Zgraded-DM-gtwined}) 
to the corresponding equivariant K3 sigma model elliptic genus, 
\begin{gather}
\label{eqn:eqK3EG}
Z_{\rm EG}(\tau,z;\Pi,g) 
:= {\rm Tr}_{\rm RR}(g (-1)^{F+\bar{F}} q^{L(0) - \frac{c}{24}} \bar{q}^{\bar{L}(0)-\frac{\bar{c}}{24}}y^{J(0)})
\end{gather}
(cf. \cite{Volpato}).
It has been checked \cite{DuncanK3} that 
\begin{gather}\label{vsnEGK3-gtwined}
Z^{s\natural}_g(\tau,z)=Z_{\rm EG}(\tau,z;\Pi,g),
\end{gather}
in all cases that a computation of the latter is available. Note that (\ref{eqn:eqK3EG}) is hard to compute in general, since the Hilbert spaces underlying K3 sigma models are, for the most part, not yet understood. By contrast, the $Z^{s\natural}_g(\tau,z)$ have been computed explicitly and uniformly in \cite{DuncanK3}. (We will recall the formula momentarily, cf. (\ref{Zint-gDM}), (\ref{Zint-gM5}).)

In this work we propose to consider a natural refinement of (\ref{Zgraded-DM}). Namely, for $g\in G_\Pi$ we define
\begin{gather}\label{Zgzw}
	Z^{s\natural}_g(\tau,z,w)
	:= {\rm Tr}(g (-1)^F q^{L(0) - \frac{c}{24}} y^{J(0)}u^{K(0)}|\vsn_\tw),
\end{gather}
where $u=e^{2\pi i w}$, and $K(z)=2K^3(z)$ for $K^3(z)$ the spin-$1$ field corresponding to
\begin{gather}\label{eqn:kappa3}
\kappa^3:=\frac14(-a_X^-(-\tfrac12)a_X^+(-\tfrac12)\vv+a_Z^-(-\tfrac12)a_Z^+(-\tfrac12)\vv)
\end{gather}
(cf. (\ref{eqn:N4c6gens})).
Then we have
\begin{equation}
\label{Zint}
Z^{s\natural}(\tau,z,w) = \frac{1}{2} \frac{1}{\eta^{12}(\tau)} \sum_{i=2}^4 (-1)^{i+1} \theta_i(\tau,z-w) \theta_i(\tau, z+w) \theta_i^{10}(\tau,0)
\end{equation}
for $Z^{s\natural}(\tau,z,w):=Z^{s\natural}_e(\tau,z,w)$. More generally, for $g\in G_{\Pi}$ we have
\begin{gather}
\begin{split}\label{Zint-gDM}
	Z^{s\natural}_g(\tau,z,w)
= 
&+\frac12\frac{\theta_1(\tau,z-w)\theta_1(\tau,z+w)}{\eta^6(\tau)}D_g\eta_g(\tau)
-\frac12 \frac{\theta_2(\tau,z-w)\theta_2(\tau,z+w)}{\theta_2^2(\tau,0)}C_{-g}\eta_{-g}(\tau)\\
&+\frac12 \frac{\theta_3(\tau,z-w)\theta_3(\tau,z+w)}{\theta_3^2(\tau,0)}\frac{\eta_{-g}(\tau/2)}{\eta_{-g}(\tau)}
-\frac12 \frac{\theta_4(\tau,z-w)\theta_4(\tau,z+w)}{\theta_4^2(\tau,0)}\frac{\eta_{g}(\tau/2)}{\eta_{g}(\tau)},
\end{split}
\end{gather}
for certain constants $C_g$ and $D_g$, defined in \cite{DuncanK3}, where $\eta_g(\tau):=\prod_{k>0} \eta(k\tau)^{m_k}$ in case the characteristic polynomial of $g$ as an operator on ${\bf 24}$ is $\prod_{k>0}(x^k-1)^{m_k}$. (E.g., $\eta_e(\tau)=\Delta(\tau)$ and $\eta_{-e}(\tau)=\Delta(2\tau)/\Delta(\tau)$, \&c.) The identity (\ref{Zint-gDM}) follows easily from Proposition 9.2 in \cite{DuncanK3}. 
Note that we can also write
\begin{gather}
\label{Zint-gM5}
Z^{s\natural}_g(\tau,z,w) 
=\frac{1}{2}\frac{1}{\eta(\t)^{12}} \sum_{i=1}^4 (-1)^{i+1} \epsilon_{g,i} \theta_i(\tau,z-w) \theta_i(\tau, z+w)\prod_{k=2}^{12} \theta_i(\t,\rho_{g,k}) 
\end{gather}
for suitable $\rho_{g,k}\in\CC$, and $\epsilon_{g,i}=\pm 1$, cf. \cite{M5}. It follows from (\ref{Zint-gDM}) or (\ref{Zint-gM5}) that 
we have
\be\label{limit_near_cusp}
\lim_{\t\to i\infty} Z^{s\natural}_g(\t,z,w) = uy+ u^{-1}y+uy^{-1} + u^{-1}y^{-1} -4+ \Tr(g|{\bf 24}).
\ee
The constant $D_g$ vanishes unless the subspace of ${\bf 24}$ fixed by $g$ is precisely $4$-dimensional (cf. (9.13) in \cite{DuncanK3}), and this explains the absence of $i=1$ terms in (\ref{Rpart}), (\ref{Zgraded-DM}), and (\ref{Zint}).

By construction, we recover (\ref{Zgraded-DM-gtwined}) from (\ref{Zgzw}) in the limit as $w$ tends to zero, 
\begin{gather}
Z^{s\natural}_g(\tau,z)=Z^{s\natural}_g(\tau,z,0).
\end{gather} 
So (\ref{Zint-gDM}) and (\ref{Zint-gM5}) also furnish explicit expressions for the $Z_g^{s\natural}(\tau,z)$. Another interesting limit is obtained when $w$ and $z$ coincide. Namely, we have
\begin{gather}
\label{Zgraded}
Z^{s\natural}(\tau,z,z) 
= \frac{1}{2} \frac{1}{\eta^{12}(\tau)} \sum_{i=2}^4 (-1)^{i+1} \theta_i(\tau,2z) \theta_i^{11}(\tau,0),
\end{gather}
which is the weak Jacobi form of weight $0$ and index $2$ that plays a leading role in the analysis of \cite{M5}, concerning ${\cal N}=2$ and ${\cal N}=4$ superconformal structures on $\vsn$ with $c=12$.
More generally, we have
\begin{gather}
\label{Zint-gM5-zz}
Z^{s\natural}_g(\tau,z,z) 
=\frac{1}{2}\frac{1}{\eta(\t)^{12}} \sum_{i=1}^4 (-1)^{i+1} \epsilon_{g,i} \theta_i(\tau,2z) 
\prod_{k=2}^{12} \theta_i(\t,\rho_{g,k}),
\end{gather}
where $\rho_{g,1}=0$ for $g\in G_\Pi$, which is the function denoted $Z_g(\tau,z)$ in \cite{M5}. Except that it is natural, in this limit, to consider twinings by elements in the larger group $G_Z> G_\Pi$, consisting of elements $g\in \Co_0$ that point-wise fix the $2$-space $\Span\{a_Z^\pm\}<{\bf 24}$, since 
\begin{gather}\label{eqn:j3k3}
\jmath^3+\kappa^3=\frac12 a_Z^-(-\tfrac12)a_Z^+(-\tfrac12)\vv.
\end{gather}
So $\rho_{g,1}$ can be non-trivial in general for $g\in G_Z$, and the investigation of the resulting functions $Z^{s\natural}_g(\tau,z,z)$, for various $G_Z$, is the main focus of \cite{M5}. In particular, there is a choice of $Z$ for which $G_Z$ is the sporadic simple Mathieu group $M_{23}$, and the corresponding ${\cal N}=2$ decompositions associate distinguished vector-valued mock modular forms to this group. (See \cite{M5} for more details on this.)

Given the important role (\ref{hugehog}--\ref{eqn:chiy}) of $Z_{\rm EG}(\t,z;K3)$ in the KKV formula (\ref{eqn:KKV}) for the
K3 Gopakumar--Vafa 
invariants, 
and the coincidence (\ref{eqn:vsnEGK3}), 
it is natural to ask about the role of $\vsn$ in the counting of BPS states in K3 compactifications, and about the relationship between symmetry group actions in these two, a priori, distinct settings. 
In the last section we will comment on conjectural positive answers to these questions. 
Furthermore, we will argue that the refined index $Z^{s\natural}(\tau,z,w)$ is a manifestation of a directly similar relationship between $\vsn$ and the refined counting of BPS states on K3.

\section{Conjectures}
\label{sec:Symmetries and Twinings}

The intimate relation 
(\ref{hugehog}--\ref{eqn:KKV}) 
between the K3 elliptic genus 
and the K3 Gopakumar--Vafa invariants on the one hand, 
and the coincidences (\ref{eqn:vsnEGK3}) and (\ref{vsnEGK3-gtwined}), relating the K3 elliptic genus to 
$\vsn$ on the other hand,
naturally lead us to the expectation that $V^{s\natural}$ 
may serve as an auxiliary BPS Hilbert space, encoding the stringy symmetries of K3 surfaces. 
In this section we 
formulate 
some 
consequences of this 
fundamental 
idea in terms of concrete and 
falsifiable conjectures on K3 surface geometry.

Our first conjecture concerns the role of the Conway group in controlling the 
symmetries that underly the 
refined K3 Gopakumar--Vafa invariants. From the discussion of the twined graded partition functions $Z_g^{s\natural}$ in \S\ref{sec:The Supernatural CFT} we can anticipate that $4$-dimensional subspaces of ${\bf 24}$ will play an important role.

As we have reviewed in \S\ref{sec:K3rev}, the reduced refined Gopakumar--Vafa invariants of a projective complex K3 surface $X$ should be realised as topological invariants of a certain D-brane moduli space, which in turn depends upon a choice of K\"ahler class and B-field.

\vspace{7pt}\noindent
{\bf Conjecture 1.}
{\em Let $X$ be a smooth projective complex K3 surface, let $J$ be a K\"ahler class for $X$, and let 
$B\in H^{1,1}(X)\cap \wtl{H}(X,\ZZ)\otimes_\ZZ\RR$. Then the supersymmetry preserving automorphism group of the D-brane moduli space that defines the corresponding reduced refined Gopakumar--Vafa invariants is $G_\Pi$ (cf. (\ref{eqn:defnGPi})), where $\Pi=\Pi_{X,Z}$ for $Z=\exp(B+iJ)$ (cf. (\ref{eqn:defnPi})). 
}
\vspace{7pt}

Recall that  $G_\Pi$ is isomorphic to a subgroup of $\Co_0$ that fixes a rank $4$ sublattice of the Leech lattice, according to the discussion in \S\ref{sec:The Supernatural CFT}. The main result of \cite{Huybrechts:2013iwa} shows that every such subgroup of $\Co_0$ arises as $G_\Pi$ for some D-brane moduli space as in the statement of Conjecture 1. Note also that $G_\Pi$ has been identified as the supersymmetry preserving automorphism group of the K3 sigma model determined by $\Pi$ in \cite{Volpato}.

Conjecture 1 
motivates us to consider the equivariant 
refined K3 Gopakumar--Vafa invariants corresponding to elements $g\in G_\Pi$. We next formulate a conjecture which predicts exactly what these 
invariants will be.
Our starting point is the relation between the KKV formula (\ref{eqn:KKV}) and the generating function (\ref{eqn:chiy}) for the $\chi_y$ genera of Hilbert schemes of K3 surfaces, as derived in \cite{MR1481482}.  The action of $G_\Pi$ on the cohomology of the K3 surface $X$ induces an action 
on the cohomology of the $n$-th symmetric product $X^{(n)}=X^n/S_n$ for each $N$. 
It is then natural to propose that the twined generating function for $\chi_y$ genera will satisfy
\be\label{eqn:chiy-gtwined}
\sum_{n\geq 0} \chi_{-y}(\Pi^{(n)},g)y^{-n} q^{n-1}  = 
(-y+2-y^{-1})\frac{\eta^6(\tau)}{\theta_1^2(\tau,z)}\frac{1}{\eta_g(\tau)}
\ee
where $\Pi^{(n)}$ denotes the sigma model on $X^{(n)}$ naturally induced from $\Pi$.

The KKV formula (\ref{eqn:KKV}) relates the Hirzebruch genera of Hilbert schemes of K3 surfaces to the 
K3 Gopakumar--Vafa invariants. We now conjecture a directly similar curve-counting interpretation for the equivariant Hirzebruch genera $\chi_{y}(\Pi^{(n)},g)$.

\vspace{7pt}\noindent
{\bf Conjecture 2.}
{\em Given $X$, $J$ and $B$ as in Conjecture 1, 
and a supersymmetry preserving automorphism $g\in G_\Pi$ of the corresponding D-brane moduli space, the $g$-equivariant reduced Gopakumar--Vafa invariants $n_{n,g}^r$ are determined by 
 \begin{equation}\label{eqn:twinedKKV}
\sum_{r\geq 0} \sum_{n\geq 0} (-1)^{(r-1)} n_{n,g}^r (y^{1/2}-y^{-1/2})^{2(r-1)} q^{n-1} = 
\frac{\eta^6(\tau)}{\theta_1^2(\tau,z)}\frac{1}{\eta_g(\tau)}
\end{equation}
where $\eta_g$ is as in (\ref{Zint-gDM}).}
\vspace{7pt}

In comparing (\ref{eqn:twinedKKV}) with (\ref{eqn:chiy-gtwined}) note that $(-y+2-y^{-1})=(-1)(y^{1/2}-y^{-1/2})^2$.

Finally, observing that the refinement from $n_{n}^r$ to $N_n^{j_L,j_R}$ is directly parallel to the refinement passing from the $\chi_y$-genus of a K3 surface to 
its Hodge numbers, we extend Conjecture 2 to a curve-counting interpretation for the generating function of the twined Hodge numbers of
$K3^{(n)}$.  In advance of the formulation, we define 
\begin{gather}\label{eqn:defnj1j2x1x2}
[j_1,j_2]_{x_1,x_2}:=\frac{(x_1^{2j_1+1}-x_1^{-2j_1-1})(x_2^{2j_2+1}-x_2^{-2j_2-1})}{(x_1-x_1^{-1})(x_2^2-x_2^{-2})-(x_2-x_2^{-1})(x_1^2-x_1^{-2})}
\end{gather}
for $j_1,j_2\in \frac12\ZZ$. To put this definition in context, note that $[j]_x=(x^{2j+1}-x^{-2j-1})/(x-x^{-1})$, and the denominator of (\ref{eqn:defnj1j2x1x2}) is the product of $(x_1-x_1^{-1})(x_2-x_2^{-1})$ and $(x_2-x_1-x_1^{-1}-x_2^{-1})$.

\vspace{7pt}\noindent
{\bf Conjecture 3.}
 {\em Given $X$, $J$ and $B$ as in Conjecture 1, and a supersymmetry preserving automorphism $g\in G_\Pi$ of the corresponding D-brane moduli space, the $g$-equivariant reduced refined Gopakumar--Vafa invariants $N_{n,g}^{j_L,j_R}$ are determined by 
\begin{gather}
\sum_{n\geq 0} \sum_{j_L,j_R \in \frac{1}{2}\ZZ_{\geq 0}}N_{n,g}^{j_L,j_R} [j_L,j_R]_{y,u} q^{n-1} \label{eqn:twinedKKV_refined}
=
\frac{\eta^6(\tau)}{\theta_1(\tau,z+w)\theta_1(\tau,z-w)}\frac{1}{\eta_g(\tau)}.
\end{gather} 
where $\eta_g$ is as in (\ref{Zint-gDM}).}
\vspace{7pt}

\section{Discussion}\label{sec:disc}

In the bulk of this paper, we focused on the relationship between K3 enumerative invariants and the 
cohomology of the symmetric products $K3^{(n)}$.  
In fact, these formulae arise from limits of the  
expression in \cite{DMVV}, relating the free energy of second quantised strings on K3 surfaces to appropriate actions of Hecke operators on $Z_{\rm EG}$,
\be\label{DMVV_K3}
\sum_{n\geq 0}Z_{\rm EG}(\t,z;K3^{[n]})p^{n-1}  = \frac{1}{p} \exp\left(\sum_{m>0} p^m (Z_{\rm EG}|V_m)(\t,z) \right), 
\ee
where the $m$-th Hecke operator $V_m$ acts on a weak Jacobi form $\f$ 
of weight $0$ according to
\be\label{eqn:DMVV}
(\f|V_m)(\t,z):=\frac{1}{m}\sum_{\substack{a,d>0\\ad=m}}\sum_{0\leq b<d}\f\left(\frac{a\t+b}{d},az\right).
\ee
(Note that $V_m$ is denoted $T_m$ in \cite{DMVV}. Our notation for Hecke operators on Jacobi forms follows \S4 of \cite{eichler_zagier}.)
Given a generator $g$ of a cyclic group $\langle g\rangle$ of order $n$, and a corresponding set of $n$ (not necessarily distinct) weak Jacobi forms $\{\f_{g^a}\}$, 
we may define an equivariant Hecke operator $V^g_m$ following \cite{MR1172696}, by setting
\be
(\f|V_m^g )(\t,z):=\frac{1}{m}\sum_{\substack{a,d>0\\ad=m}}\sum_{0\leq b<d}\f_{g^a}\left(\frac{a\t+b}{d},az\right). 
\ee

Now suppose that $X$ and $Z$ are as in 
the statement of 
Conjecture 1, and $\Pi=\Pi_{X,Z}$. We regard $\Pi$ as a sigma model with target a K3 surface $X$, 
according to the prescription of \cite{MR1479699,MR1416354,NahWen},
and write $\Pi^{(n)}$ for the corresponding sigma model on the 
$n$-th symmetric power $X^{(n)}$. 
Then a supersymmetry preserving automorphism $g$ of $\Pi$ lifts naturally to each $\Pi^{(n)}$. 
A formula for the generating function of the equivariant elliptic genera $Z_{{\rm EG}}(\t,z;\Pi^{(n)},g)$ 
has been proposed in \cite{Cheng:2010pq}. Using the $V_m^g$ it reads
\be\label{eqn:DMVV-gtwined}
\sum_{n\geq 0} Z_{{\rm EG}}(\t,z;\Pi^{(n)},g)p^{n-1}  = \frac{1}{p} \exp\left(\sum_{m>0} p^m (Z_{\rm EG}|V_m^g)(\t,z;\Pi,g) \right) . 
\ee
Alternatively, defining $c_{g}(4n-\ell^2)\in\CC$ by requiring that
\be
Z_{{\rm EG}}(\t,z;\Pi,g) = \sum_{\substack{\ell,n\in\ZZ\\n\geq 0}} c_{g}(4n-\ell^2) q^ny^\ell, 
\ee
the formula (\ref{eqn:DMVV-gtwined}) can be rewritten 
\be\label{eqn:DMVV-gtwined-alt}
\sum_{n\geq 0} Z_{{\rm EG}}(\t,z;\Pi^{(n)},g)p^{n-1}  = \frac{1}{p} \exp\left(\sum_{k>0} \frac{1}{k} \sum_{\substack{n,n',\ell\in \ZZ\\n\geq 0, n'>0}}c_{g^k} (4nn'-\ell^2) (q^np^{n'}y^\ell)^k \right) . 
\ee
The formula in Conjecture 2 of \S4\ is a special limit of this one (as $q\to 0$).  It would be natural to attach an interpretation in terms of BPS states to
(\ref{eqn:DMVV-gtwined-alt}).   
In particular, to test for a deeper connection between BPS states of K3 and an auxiliary module such as $V^{s\natural}$, probing more than just
the geometry of the Ramond ground states, it would be nice to extend the comparison to BPS states which preserve a smaller fraction of the
supersymmetry.  Similar comments would apply to purported connections between K3 geometry and modules associated to Mathieu or Umbral
moonshine.

We now discuss two further observations about Conjectures 2 and 3, and then conclude with some questions.

As we have noted, Conjecture 2 can be regarded as a limiting case of Conjecture 3. 
Our first observation concerns the other interesting limit of 
families of M-theory compactifications giving rise to refined Gopakumar--Vafa invariants. Namely, we may consider the NS limit (cf. (\ref{rel_refined_unrefinedGV})) in which one only keeps track of the quantum number $j_L+j_R$ in $N_{\b}^{j_L,j_R}$. In the present context of  K3 curve counting, the corresponding generating function is obtained by setting $u=v$ in (\ref{refined}). Using the same arguments as above, we conjecture that 
the Ramond ground states of $\vsn_\tw$ also serve as auxiliary Hilbert spaces in this limit, but with graded Ramond sector partition function obtained by setting $z=w$ in
(\ref{Zint}), and given explicitly by the weight 0 index 2 Jacobi form (\ref{Zgraded}). 
Note that in this limit 
we have 11 pairs of uncharged fermions instead of 10, which suggests that one should be able to consider twinings of (\ref{Zgraded}) by elements of $\Co_0$ 
whose fixed-point sublattice of the Leech lattice has rank as low as $2$ (as opposed to $4$).
It is hence natural to 
expect that the corresponding M-theory compactification admits an enhanced symmetry group in the NS limit. See also the discussion of $G_Z$ in \S\ref{sec:The Supernatural CFT} (cf. (\ref{eqn:j3k3})) in this regard.

Our second 
observation concerns the automorphic aspects of our conjectures. 
As we have detailed in \S\ref{sec:The Supernatural CFT}, the refined partition function 
$Z^{s\natural}(\tau,z,w)$ 
specializes to weak Jacobi forms of 
index $1$ and $2$, respectively, when the respective limits $w\to 0$ and 
$w\to z$ are taken. This is explained by the fact that $Z^{s\natural}(\tau,z,w)$ is a Jacobi form of index $1$ for the rank $2$ lattice $A_1\oplus A_1$, in the sense of \cite{MR3123592}. (Cf. Example 2.7 of \cite{MR3123592}. The usual Jacobi forms, of \cite{eichler_zagier}, are associated with the even lattices of rank $1$.) A similar statement holds for the twined counterpart $Z^{s\natural}_g(\tau,z,w)$, for $g\in G_\Pi$, where the 
modular invariance group 
is replaced by 
$\Gamma_0(N_g)$, for some $N_g\in\ZZ^+$ depending on $g$. 

Also of relevance is the fact that the generating function of the $Z_{\rm EG}(\t,z;K3^{[n]})$ is, famously, almost automorphic. 
More precisely, we can rewrite (\ref{DMVV_K3}) as 
\be\label{DMVV_K32}
 \sum_{n\geq 0} Z_{\rm EG}(\t,z;K3^{[n]})p^{n-1}  = \frac{\f_{10,1}(\s,z)}{\Phi_{10}(\t,\s,z)} ,
\ee
where $p=e^{2\p i \s}$, we write 
$\Phi_{10}$ for the 
Igusa form, being the 
(unique up to scale) 
cusp form of weight $10$ for $\Sp_4(\ZZ)$, and 
$ \f_{10,1}:=-\th_1^2\eta^{18}$ is a cuspidal Jacobi form of weight $10$ and index $1$. Note that we have 
\be\label{eqn:prodf101}
\f_{10,1}(\t,z) 
= q \left(\prod_{k>0} (1-q^k)\right)^{c(0)}   \left( (y^{1/2}-y^{-1/2}) \prod_{k>0} (1-q^ky)(1-q^k y^{-1}) \right)^{c(-1)}
\ee
where $c(0)=20$ and $c(-1)=2$ 
(cf. (\ref{eqn:c(4n-ell2)})).

Physically, the correction factors $ \frac1{\eta^{24}(\s)}$ and $-\frac{\eta^{6}(\s)}{ \theta_1^2(\s,z)}$ in $\frac{1}{\f_{10,1}}$ 
may be interpreted as arising from the CFT describing the Taub-NUT and the center of mass degrees of freedom of the D1-D5 system \cite{David:2006yn,Shih:2005uc}. The connection between $\Phi_{10}(\t,\s,z)$ and $Z_{\rm EG}(\t,z;K3)$ can be made more direct, by noting that  the former can be expressed as 
a multiplicative lift of the latter. Namely, we have
\be\label{eqn:Phi10prod}
\Phi_{10}(\t,\s,z) = pqy\prod_{\substack{r,s \geq 0, t \in \ZZ\\t<0 \;{\rm if }\; r=s=0}} (1-q^sy^tp^r)^{c(4rs-t^2)},
\ee
where the $c(4rs-t^2)$ are as in (\ref{eqn:c(4n-ell2)}).
Note also that, apart from the 
product formula (\ref{eqn:Phi10prod}), one also has an infinite sum expression for $\Phi_{10}(\t,\s,z)$, as the 
additive lift (i.e. generating function of images under Hecke operators) of the Jacobi form 
$\f_{10,1}$.

It is natural to expect that the refined Ramond sector index 
$Z^{s\natural}(\tau,z,w)$ 
enjoys an analogous 
relation, to 
an automorphic form for (a subgroup of) the orthogonal group of $U^2\oplus A_1^2(-1)$. Cf. \cite{MR3123592,2012arXiv1203.6503G}. 
We refrain for now from a general discussion, and focus on the NS limit, $Z^{s\natural}(\tau,z,z)$.
First note that  the multiplicative lift of $Z^{s\natural}(\tau,z,z)=\sum \tilde c(n,\ell)q^ny^\ell$, defined by 
\be
\D_{11}(\t,\s,z) =  qyp^2 \prod_{\substack{r,s \geq 0, t \in \ZZ\\t<0 \;{\rm if }\; r=s=0}} (1-q^sy^tp^{2r})^{\tilde c(rs,t)},
\ee
is a 
modular form of
weight $11$ for the paramodular group $\Gamma_2$, according to Example 3.4 of \cite{GriNik_AutFrmLorKMAlgs_II}. Moreover, 
$\D_{11}$  coincides (cf. also Example 3.4 of \cite{GriNik_AutFrmLorKMAlgs_II}) with the  additive lift of the weight 11 index 2 Jacobi form $\f_{11,2}(\tau,z):=-i\th_1(\tau,2z)\eta^{21}(\tau)$, which satisfies
\be
\f_{11,2}(\t,z) 
= q \left(\prod_{k>0} (1-q^k)\right)^{\til c(0,0)}    \left((y-y^{-1})\prod_{k>0}(1-q^k y^2)(1-q^k y^{-2})\right)^{\til c(0,2)}.
\ee
This may be compared to (\ref{eqn:prodf101}).
Finally, we note that  
\be
\sum_{n\geq 0}\sum_{j \in \ZZ_{\geq 0}} \n_{n}^j [j]_{y^2}  \,p^{2n-2} = 
\lim_{\t\to i\infty} \frac{\f_{11,2}(\s,z)}{\D_{11}(\t,\s,z)},
\ee
in direct analogy with (\ref{DMVV_K32}). 
In the above we write $\n_n^j$ for the NS limit invariants, defined via
\be
\sum_{j \in \frac{1}{2}\ZZ_{\geq 0}}\, \n_{n}^j [j]_{y^2} = \sum_{j_L,j_R \in \frac{1}{2}\ZZ_{\geq 0}} N_n^{j_L,j_R} [j_L]_y [j_R]_y . 
\ee
We list the first few nonzero NS limit invariants up to $n=4$ in Table \ref{tbl:NSinv}. We expect analogous relations to automorphic forms to persist for the twined refined K3 Gopakumar--Vafa invariants. 

\begin{table}
\begin{center}
\begin{tabular}[htb]{c|ccccc}
$ \n_{n}^j$&$j=0$&1/2&1&3/2&2\\\hline
 $n=0$&1&&&&\\
 1&22&1&&&\\
 2&275&23&1&&\\
 3&2531&298&23&1&\\
 4&18998&2829&299&23&1
\end{tabular}\caption{The first few nonzero invariants $\n_{n}^j$ for $n\leq 4$.}\label{tbl:NSinv}
\end{center}
\end{table}

The conjectures in the present paper raise many questions. We will close the note with a short discussion of a few of them. 
First, it would be of great geometrical interest to verify (or disprove) {Conjectures 2} and {3}, for 
symmetries $g\in G_\Pi$ that are inherited from symplectomorphisms of the underlying K3 surface. 
Note that some relevant calculations, in agreement with our conjectures, appear in \cite{2015arXiv150506420H}. Second, the above observation regarding the special feature that appears in the NS limit should be better understood.  
Third, a natural question is to which extent other K3 compactifications in string theory enjoy a similar relation to finite groups. (See for instance \cite{Cheng:2013kpa} for a preliminary exploration in the landscape of type II compactifications on K3-fibered threefolds, and their dual descriptions via K3 compactifications of heterotic strings.)
Finally, from both the physical and number theoretic points of view, it would be interesting to develop the discussion above in the setting of automorphic forms for $U^2\oplus A_1^2(-1)$, and study in more detail the interpolation 
between the distinguished forms $\Phi_{10}$ and $\D_{11}$ appearing above.

\section*{Acknowledgements}

We thank the Perimeter Institute for Theoretical Physics for hospitality during discussions of the 
physics of this note, and the participants of ``(Mock) Modularity, Moonshine, and String Theory" for creating a stimulating environment.  
We are grateful to the organisers of the LMS - EPSRC Durham Symposium on ``New moonshines, mock modular forms and string theory" for providing an ideal setting to finish this work.
M.C. thanks Sheldon Katz and Cumrun Vafa for helpful conversations, and S.H. thanks Babak Haghighat and Guglielmo Lockhart for answering questions related to the contents of this note. S.K. is grateful to the Aspen Center for Physics for hospitality during both the inception and the late stages of this work.  J.D. gratefully acknowledges support from the U.S. National Science Foundation (DMS 1203162), and from the Simons Foundation (\#316779), and thanks the University of Tokyo for hospitality during the final stages of this work. S.H. is supported by the Harvard University Golub Fellowship in the
Physical Sciences. S.K. is supported by the NSF via grant PHY-1316699 and the DoE Office of
Basic Energy Sciences contract DE-AC02-76SF00515.

\appendix

\section{Theta Functions}\label{sec:theta}

We define the Jacobi theta functions by setting
\begin{align}
\begin{split}\label{equation:theta1}
\theta_1(\tau,z) & :=-i\sum_{n\in \ZZ}(-1)^n y^{n+1/2} q^{(n+1/2)^2/2}, \\
\theta_2(\tau,z) & :=\sum_{n\in\ZZ}y^{n+1/2}q^{(n+1/2)^2/2}, \\
\theta_3(\tau,z) & :=\sum_{n\in\ZZ}y^n q^{n^2/2}, \\
\theta_4(\tau,z) & :=\sum_{n\in\ZZ}(-1)^n y^n q^{n^2/2}. \\
\end{split}
\end{align}
We have the product formulas
\begin{align}
\begin{split}\label{eqn:autfms-theta1prod}
\theta_1(\tau,z) 
& =-i q^{1/8}y^{1/2}(1-y^{-1})\prod_{n>0}(1-y^{-1}q^n)(1-yq^n)(1-q^n), \\
\theta_2(\tau,z)
& =q^{1/8}y^{1/2}(1+y^{-1})\prod_{n>0}(1+y^{-1}q^n)(1+yq^n)(1-q^n), \\
\theta_3(\tau,z) 
& =\prod_{n>0}(1+y^{-1}q^{n-1/2})(1+y q^{n-1/2})(1-q^n), \\
\theta_4(\tau,z)
& =\prod_{n>0}(1-y^{-1}q^{n-1/2})(1-yq^{n-1/2})(1-q^n), \\
\end{split}
\end{align}
by virtue of the Jacobi triple product identity.

\newpage

\end{document}